\documentstyle[12pt]{article}% 11pt, article report letter
\begin{document}
\vspace*{-1in}
\begin{flushright}
CERN-TH.7329/94 \\
TIFR/TH/94-21 \\
June 1994
\end{flushright}
\vskip 65pt
\begin{center}
{\Large \bf \boldmath Fragmentation contribution to quarkonium
production in hadron collision}\\
\vspace{8mm}
{D.P.~Roy$^{\dagger}$}\\
\vspace{5pt}
{\it Theory Group, Tata Institute of Fundamental Research,\\
Homi Bhabha Road, Bombay 400 005, India.}\\
\vspace{15pt}
{and}\\
\vspace{12pt}
{K. Sridhar$^{*}$}\\
\vspace{5pt}
{\it Theory Division, CERN, \\ CH-1211, Geneva 23, Switzerland.}\\

\pretolerance=10000

\vspace{70pt}
{\bf ABSTRACT}
\end{center}
We compute the contributions due to gluon and heavy-quark
fragmentation to quarkonium production at large transverse
momentum in $\bar p p$ and $p p$ collisions.
For inclusive $J/\psi$ production, there is a large
contribution from the $g \rightarrow \chi_c$ fragmentation.
At large $p_T$, this is comparable to the conventional charmonium
model prediction via gluon fusion at the ISR and dominates over
the latter at the Tevatron energy. This may help to explain, at
least partly, the large $J/\psi$ production cross-section recently
observed by the CDF experiment. However, the
fragmentation contribution to $\psi^{\prime}$ production
is not large enough to explain the corresponding CDF data.
We also present the results
for $\Upsilon$ production at ISR and Tevatron energies.
\vspace{40pt}
\noindent
\begin{flushleft}
CERN-TH.7329/94\\
June 1994\\
\vspace{11pt}
$^{\dagger)}$ dproy@theory.tifr.res.in \\
$^{*)}$ sridhar@vxcern.cern.ch\\
\end{flushleft}

\vfill
\clearpage
\setcounter{page}{1}
\pagestyle{plain}
The production of quarkonia in hadron-hadron and lepton-hadron
collisions has been the subject of several
theoretical and experimental investigations.
Experimentally, the leptonic decay channels of quarkonia provide
the cleanest signals of heavy quark production. From the
theoretical point of view, quarkonium production
provides a very important test of perturbative
QCD. Both in lepton-hadron and in hadron-hadron collisions,
the quarkonia are produced mainly through gluon-initiated
subprocesses~-- i.e. photon-gluon and gluon-gluon fusion leading to
the heavy-quark pair production. Consequently these processes are
also important probes for the gluon distribution in the hadron.

The colour-singlet model has been used, with reasonable
success, to describe the leptoproduction \cite{berjon}
and hadroproduction \cite{br} of quarkonia at large transverse
momentum. In this QCD-based approach, one projects out from the
full heavy-quark pair production amplitude, the part with the correct
spin, parity and charge-conjugation assignments of a ${}^{2S+1}L_J$
quarkonium state. The $Q \bar Q$ is required to be a colour singlet,
which is achieved by the radiation of a hard gluon in the final
state. The matrix-element so obtained is then convoluted with
the wave-function at the origin, $\vert R_0 \vert^2$ (or its
derivative, in the case of $P$-wave quarkonia). In this approach,
it is possible to compute the production cross-sections for the
different resonances in a given family of quarkonia: in particular,
the $P$-wave cross-sections are computable and their contribution to
the inclusive cross-section of the $S$-state completely specified,
once the branching ratios of the $P$-states to the $S$-state are
determined experimentally. The colour-singlet model is known
to provide a reasonable description of data from the EMC and
NMC experiments on large-$p_T$ leptoproduction \cite{emc,nmc}
of $J/\psi$ and on large-$p_T$ hadroproduction \cite{br,gms} of
$J/\psi$ from ISR experiments. For a recent review of quarkonium
production, and for a discussion of the $J/\psi$ production data
we refer the reader to Ref.~\cite{schuler}.

The recent CDF data on large-$p_T\ J/\psi$ production in 1.8~TeV
$\bar p p$ collisions at the Tevatron \cite{cdf}
seem to indicate, however, a serious discrepancy~:
the theoretical
predictions of the above charmonium model
are well below the data, which suggests the existence
of a new contribution to $J/\psi$ production at large $p_T$.
It has been pointed out by Braaten and Yuan \cite{bryu} that a
large contribution to quarkonium production at large $p_T$
could come from the fragmentation of gluons and heavy
quarks. Even though the fragmentation process is of higher order
in the strong coupling constant, $\alpha_s$, it can dominate
over the direct quarkonium production via fusion at large
$p_T$, where terms of the order of $p_T^2/m_Q^2$
can easily compensate for the
suppression due to the extra power of $\alpha_s$.
It is important to check whether the fragmentation process
can explain the discrepancy between the theoretical predictions for
$J/\psi$ production and the data from CDF.

While it is important to study the fragmentation contribution
at the Tevatron energy in the light of the new CDF data, it is
also interesting to study the magnitude of this contribution at
lower energies. This is particularly important in view of the
fact that the low-energy $J/\psi$ production data has served
as a benchmark for signals of quark-gluon plasma (QGP) formation
such as $J/\psi$ suppression \cite{matsa}. Indeed, such a
suppression of $J/\psi$ production has been observed \cite{na38}
by the NA38 experiment in heavy-ion collisions at SPS energies
($\sqrt{s}\approx 20$~GeV). One important feature
of the QGP signal is the $p_T$ dependence of the suppression; therefore,
an estimate of the magnitude of the fragmentation contribution
to $J/\psi$ production at these energies is important.

In this letter, we present a complete perturbative QCD calculation of
$J/\psi$ production at Tevatron energy, which includes the
colour-singlet model prediction for the fusion contribution as well as
the fragmentation contribution. We find that the gluon fragmentation
contribution is large and can help at least partially to resolve the
discrepancy between the CDF data and the earlier theoretical
predictions. We also study these contributions for $pp$ collisions
at $\sqrt{s}=63$~GeV and compare them to the data from the ISR
experiment \cite{isr}. Results for $\psi^{\prime}$
production at Tevatron and at ISR energies are also presented.
In this case the fragmentation contribution is not large enough
to explain the discrepancy between the fusion contribution and
the CDF data at the Tevatron energy. This may indicate a sizable
nonperturbative contribution to charmonium production, presumably
due to the small charm quark mass. Therefore, we present the
analogous results for $\Upsilon$ production, which may be
compared with the future $\Upsilon$ production data from the
Tevatron.

After this work was completed, results for $J/\psi$
production at Tevatron energy were
presented in Refs.~\cite{cg,bdfm}. Ref.~\cite{bdfm} has also
presented results for $\psi^{\prime}$ production at Tevatron
energy. Our results for $J/\psi$ and $\psi^{\prime}$ production
at the Tevatron energy are in agreement with the results
in Refs.~\cite{cg,bdfm}. However, as emphasised above, we have also
studied the energy dependence of the fragmentation contribution~-- in
particular, we have studied $J/\psi$ and $\psi^{\prime}$
production at ISR energies. We also present results for
$\Upsilon$ production at ISR and Tevatron energies.

In computing the inclusive $J/\psi$ $p_T$ distributions, for
the fusion and the fragmentation processes, the contributions
of the $P$-wave $\chi_c$ states have also to be considered. The
$\chi$'s are produced copiously and the $\chi_1$ and $\chi_2$
have a sizeable branching into the $J/\psi + \gamma$ mode.
The large-$p_T$ production cross-section for the fusion process is
given as
\begin{eqnarray}
\label{e1}
&&E{d\sigma \over dp^3}(AB \rightarrow (J/\psi,\chi_i) X)
=  \nonumber \\
&& \sum \int dx_1 x_1G_{a/A}(x_1) x_2G_{b/B}(x_2)
{2 \over \pi} {1 \over 2x_1 -\bar x_T e^y}
{d\hat \sigma \over d \hat t}(ab \rightarrow
(J/\psi,\chi_i) c) .
\end{eqnarray}
In the above expression, the sum runs over all the partons
contributing to the subprocesses $ab \rightarrow (J/\psi,\chi_i) c$;
$G_{a/A}$ and $G_{b/B}$ are the distributions of the partons $a$ and $b$
in the hadrons $A$ and $B$ with momentum fractions $x_1$ and
$x_2$, respectively. Energy-momentum conservation determines
$x_2$ to be
\begin{equation}
\label{e2}
x_2= {x_1 \bar x_T e^{-y} - 2 \tau \over 2x_1-\bar x_T e^y},
\end{equation}
where $\tau = M^2/s$, with $M$ the mass of the resonance, $s$
the centre-of-mass energy and $y$ the rapidity at which the resonance
is produced.
\begin{equation}
\label{e3}
\bar x_T= \sqrt{x_T^2 + 4\tau} \equiv {2M_T \over \sqrt{s}},
\hskip20pt x_T={2p_T \over \sqrt{s}}
\end{equation}
The expressions for the subprocess cross-sections, $d\hat\sigma/d\hat t$,
are explicitly given in Refs.~\cite{br} and \cite{gtw}. We obtain
$d\sigma /dp_T$ from Eq.~\ref{e1} using
\begin{equation}
E{d\sigma \over dp^3} = {1\over \pi} {1 \over 2p_T}
{d\sigma \over dy dp_T}~.
\end{equation}

The fragmentation contribution is computed by factorising the
cross-section for the process $AB \rightarrow (J/\psi,\chi_i) X$ into a
part containing the hard-scattering cross-section for producing a
gluon or a charm quark and a part which specifies the fragmentation of
the gluon (or the charm quark) into the required charmonium state, i.e.
\begin{equation}
\label{e4}
d\sigma (AB \rightarrow (J/\psi,\chi_i) X)
 = \sum \int_0^1 dz \hskip4pt
d\sigma (AB \rightarrow c X) D_{c \rightarrow (J/\psi,\chi_i)}(z,\mu ) ,
\end{equation}
where $c$ is the fragmenting parton (either a gluon or a charm quark)
and the sum in the above equation runs over all contributing partons.
$D(z,\mu)$ is the fragmentation function and $z$, as usual, is
the fraction of the momentum of the parent parton carried by the
charmonium state. The fragmentation function is
computed perturbatively at an initial scale $\mu_0$
which is of the order of $m_c$. It is then evolved to the scale
typical of the fragmenting parton which is of the order of
$p_T/z$, using the Altarelli-Parisi equation:
\begin{equation}
\label{e5}
\mu {\partial \over \partial\mu} D_{i\rightarrow (J/\psi,\chi_i)} (z)
= \sum_j\int_z^1{dy \over y} P_{ij}({z\over y}, \mu)
D_{j\rightarrow (J/\psi,\chi_i)}(y) ,
\end{equation}
where the $P_{ij}$ are the splitting functions of a parton $j$
into a parton $i$. We consider the fragmentation of gluons and
charm quarks alone.
In principle, the contribution of the light quarks should also be
considered but their contribution to the fragmentation at large $p_T$
is small and can be neglected. Further, in the evolution we
consider only the $P_{gg}$ contribution in evolving the gluon
fragmentation function and the $P_{cc}$ contribution in evolving
the charm quark fragmentation. The effect of the
non-diagonal splitting function contributions can be safely neglected.
The full set of initial fragmentation functions that we need to
obtain the $J/\psi$ and the $\chi$ contributions have now been
computed. These are $D_{g \rightarrow J/\psi}$ \cite{bryu},
$D_{g \rightarrow \chi}$ \cite{bryu2}, $D_{c \rightarrow \psi}$
\cite{brcyu} and $D_{c \rightarrow \chi}$ \cite{chen,yuan}.

For the fragmentation process, the cross-section is given by a
formula similar to Eq.~\ref{e1} but with an extra integration
over $z$, or equivalently over $x_2$. We have
\begin{eqnarray}
\label{e6}
&&E{d\sigma \over dp^3}(AB \rightarrow (J/\psi,\chi_i) X)
=  \nonumber \\
&& \sum \int dx_1 dx_2 G_{a/A}(x_1) G_{b/B}(x_2)
D_{c\rightarrow (J/\psi,\chi_i)} (z)
{1 \over \pi z} {d\hat \sigma \over d \hat t}(ab \rightarrow cd) ,
\end{eqnarray}
with $z$ given by
\begin{equation}
z= {\bar x_T \over 2} ({e^{-y} \over x_2} + {e^y \over x_1}) .
\end{equation}
For $d\hat\sigma/d\hat t(ab \rightarrow cd)$, we have used the
lowest-order expressions.

Using the formalism described above, we can compute the direct
(fusion) and
fragmentation contributions to $J/\psi$ production. To do so, we
have to choose a set of parton distributions that are compatible
with all the available information on structure functions.
For $J/\psi$ production at ISR energies this is not very crucial;
however, at Tevatron energies the values of $x$ probed are
small, and one should use a set of parton densities which are
compatible with the low-$x$ structure functions measured at HERA.
In our computations, we have used \cite{plothow} the updated
MRSD-${}^{\prime}$ parametrisations \cite{mrs} for the parton
densities in the nucleon, which are compatible
with the most recent data on structure functions from HERA.
The parton densities are evolved to a scale $Q^2=\mu^2/4$, where
$\mu$ is chosen to be $M_T$ for the case of direct
$J/\psi$ production, and equal to $p_T^{g,c}=p_T/z$ for $J/\psi$
production $via$ fragmentation. The fragmentation functions are
evolved to the scale $p_T/z$. We use the above choice of
parameters for all the results presented below, except when we
vary these parameters to see the sensitivity of our results to
this choice. Another uncertainty that enters the normalisation of the
cross-section predictions is that due to the wave-function at the
origin, $R_0$, and the derivative for the $P$-states, $R_1^{\prime}$.
In the computation of the fusion contribution these appear in the
subprocess cross-sections, whereas for the fragmentation contribution,
the fragmentation functions at the initial scale are proportional to
these wave-function factors. In Ref.~\cite{bryu2} the fragmentation
function for $g\rightarrow \chi$ is written in terms of two parameters
$H_1$ and $H_8^{\prime}$, where $H_1$ is related to
$R_1^{\prime}$, and $H_8^{\prime}$ is a parameter that describes the
$g \rightarrow \chi$ fragmentation $via$ a colour-octet mechanism. The
parameter $H_8^{\prime}$ is rather poorly determined. For the parameters
$R_0$, $H_1$ and $H_8^{\prime}$, we have used the values quoted in
Refs.~\cite{bryu,bryu2} ($R_0^2=0.8$~GeV${}^3$, $H_1=15.0$~MeV,
$H_8^{\prime}=3.0$~MeV).

The results for inclusive $J/\psi$ production for both the fusion and
fragmentation contributions at ISR ($\sqrt{s}=63$~GeV) and Tevatron
($\sqrt{s}=1.8$~TeV) have been shown in Fig.~1. To be able to compare
directly with the data from $p p$ collisions at the ISR, we have
computed the invariant cross-section $BEd\sigma/dp^3$ at $y=0$
as a function of $p_T$, where $B$ is the $J/\psi$ branching ratio
into leptons ($B=0.0594$). For $\bar p p$ collisions at the Tevatron,
we present results for $Bd\sigma/dp_T$ integrated over a pseudo-rapidity
range $\vert \eta \vert < 0.5$.
At the Tevatron energy, the $b$-quark contribution to $J/\psi$
production is substantial, but this contribution has been removed
from the data by the use of a micro-vertex detector in the CDF
experiment \cite{cdf}.
The fusion contribution (including both $J/\psi$ and $\chi$
contributions, shown as the dashed line in Fig.~1) is significantly
below the Tevatron data. At this energy, the gluon fragmentation
contribution (again, including $g \rightarrow
J/\psi$ and $g \rightarrow \chi$ and shown as the dashed-dotted
line in Fig.~1) is the dominant contribution over almost the
whole range of $p_T$ values considered.
The charm-quark fragmentation (shown as the dotted line) is
insignificantly small. The sum of all three contributions
(which is shown as the full line) is consistent with the data
within a factor of $\sim 2$. The comparison with the ISR data
also reveals some interesting features. The gluon fragmentation
contribution is now smaller than half the fusion contribution at
the lowest values of $p_T$ considered, but grows steadily with
increasing $p_T$. Addition of the fragmentation component is
seen to improve the agreement with the ISR data significantly.

In Fig.~2, we show the magnitude of the variation of the total
contribution, due to changing the scale from $\mu/2$ to $2\mu$.
These correspond to the upper and the lower lines respectively.
It should be mentioned here that the former choice is favoured
by several QCD tests.
We have also studied the effect of varying the choice of parton
densities by using GRV densities \cite{grv}
instead of MRSD-${}^{\prime}$,
since both are compatible with the most recent
structure function measurements. The effect of this change on
the total $J/\psi$ cross-section is not significant.

Our results for $\psi^{\prime}$ are shown in Fig.~3, again for
both the energies. For $\psi^{\prime}$, the $P$-state contributions
are, of course, absent.
At the lower ISR energy, the fragmentation contributions
are much too small so that the total contribution is equal to the
direct contribution. For the Tevatron energy, we find that the charm
quark fragmentation contribution dominates at large $p_T$.
The $\psi^{\prime}$ data is also contaminated at this energy
by the $b$-decay contributions. We use the $b$-inclusive
data from CDF and subtract the theoretical $b$-quark contribution
to get the data points shown in Fig.~3.
But the sum of all the contributions is clearly not anywhere
close to explaining the data. This is clearly an indication
that some new mechanism is needed to explain the $\psi^{\prime}$
data. It may be relevant to note here that the $\psi^{\prime}$
has only the $S$-wave contribution, whereas the $\psi$ is dominantly
produced from the decays of the $P$-wave $\chi$ states. A significant
contribution from any new mechanism to $S$-state production will
therefore lead to a relatively larger enhancement for
$\psi^{\prime}$ compared to $\psi$. In view of the modest value of
the charm quark mass one may speculate such a contribution to
arise from nonperturbative effects.

Fig.~4 shows our predictions for the case of $\Upsilon$ production.
As in the case of $J/\psi$ production, we have included the
contributions due to the decays of the $P$-states.
At the ISR energy, the $\Upsilon$ comes largely from direct production
via fusion while the fragmentation contributions are negligibly
small. For the Tevatron energy, the gluon fragmentation contribution
starts dominating only beyond a $p_T$ of 15~GeV; the relatively
larger value of this cross-over point reflects the relatively
larger mass of bottom quark compared to charm. This also implies that
any nonperturbative contribution will be relatively small for the
$\Upsilon$. Thus it will be important to test this perturbative QCD
prediction with the $\Upsilon$ production data from the Tevatron.

To conclude, we have studied both the fusion and fragmentation
contributions to $J/\psi$, $\psi^{\prime}$ and $\Upsilon$ production
at Tevatron and ISR energies. We find that at the Tevatron energy,
the $J/\psi$'s are produced dominantly $via$ gluon fragmentation.
The sum of the $J/\psi$ production cross-sections from the fusion
and fragmentation processes is compatible with the data from CDF
to within a factor of $\sim 2$. For $\psi^{\prime}$, however, even the
addition of the fragmentation contribution is not sufficient to
restore agreement with the CDF data. This may indicate a significant
nonperturbative contribution to charmonium production, presumably
due to the modest value of the charm quark mass. Therefore it will
be important to test the predicted $\Upsilon$ production cross-section
with the future Tevatron data.

We thank Drs. G. Altarelli and M. Mangano for several helpful
discussions. One of us (DPR) acknowledges the hospitality of the
Theory Division, CERN, during the course of this work.

\newpage
\section*{Figure captions}
\renewcommand{\labelenumi}{Fig. \arabic{enumi}}
\begin{enumerate}
\item
Upper figure: The cross-section $BEd\sigma /dp^3$ (at $y=0$) for
the process $pp \rightarrow J/\psi X$ as a function of $p_T$ at
$\sqrt{s}=63$~GeV. The data are taken from Ref.~\cite{isr}.
Lower figure: The cross-section $Bd\sigma/dp_T$ (integrated over the
pseudorapidity range $-0.5< \eta < 0.5$) for the process
$\bar p p \rightarrow J/\psi X$ as a function of $p_T$ at
$\sqrt{s}=1.8$~TeV. The data are taken from Ref.~\cite{cdf}.
The different curves (in both the upper and the lower figures)
correspond to the direct production via fusion (dashed line), the
gluon fragmentation contribution (dashed-dotted line), the charm
quark fragmentation term (dotted line) and the sum of all contributions
(solid line).

\item
The scale dependence of the $J/\psi$ cross-section as a function of $p_T$
for $pp$ collisions at $\sqrt{s}=63$~GeV (upper figure), and for
$\bar p p$ collisions at $\sqrt{s}=1.8$~TeV (lower figure). The solid
curve in both figures is for the scale $\mu/2$ and the dashed
curve is for the scale $2\mu$.

\item
The $\psi^{\prime}$ cross-section as a function of $p_T$ for
$pp$ collisions at $\sqrt{s}=63$~GeV (upper figure), and for
$\bar p p$ collisions at $\sqrt{s}=1.8$~TeV (lower figure),
with the different contributions shown as in Fig.~1. The
$\sqrt{s}=1.8$~TeV data are obtained as explained in the text.

\item
The $\Upsilon$ cross-section as a function of $p_T$ for
$pp$ collisions at $\sqrt{s}=63$~GeV (upper figure), and for
$\bar p p$ collisions at $\sqrt{s}=1.8$~TeV (lower figure),
with the different contributions shown as in Fig.~1. The
$\sqrt{s}=63$~GeV data are taken from Ref.~\cite{isr}.

\end{enumerate}
\end{document}